\documentclass[a4paper,twoside]{article}

\usepackage{subfigure}
\usepackage{calc}
\usepackage{amssymb}
\usepackage{amstext}
\usepackage{amsmath}
\usepackage{amsthm}
\usepackage{multicol}
\usepackage{pslatex}
\usepackage{apalike}
\usepackage{SciTePress}
\usepackage{graphics}
\usepackage{graphicx}
\usepackage[small]{caption}
\usepackage[all,knot,poly,dvips]{xy}

\begin{document}

\title{\uppercase{Fault-tolerant mosaic encoding 
in knot-based cryptography}  
}
\author{\authorname{Annalisa Marzuoli\sup{1,3} and Giandomenico Palumbo\sup{2,3} }
\affiliation{\sup{1} Dipartimento di Matematica `F. Casorati',
Universit\`a degli Studi di Pavia, via Ferrata 1,
27100 Pavia, Italy}
\affiliation{\sup{2}Dipartimento di Fisica,
Universit\`a degli Studi di Pavia, 
via A. Bassi 6, 27100 Pavia, Italy}
\affiliation{\sup{3}Istituto Nazionale di Fisica Nucleare, Sezione di Pavia\\
via A. Bassi 6, 27100 Pavia, Italy}
\email{\{annalisa.marzuoli, giandomenico.palumbo\}@pv.infn.it}
}


\abstract{
The  cryptographic
protocol based on topological knot theory,
recently proposed by the authors, 
is improved for what concerns the efficiency of the
encoding of  knot diagrams and its  error robustness. 
The standard Dowker--Thistlethwaite code,
based on the ordered assignment of two numbers to each crossing
of a knot diagram and not unique for some classes of knots,
is replaced by a system of eight  prototiles (knot mosaics)
which, once assembled according to a set of combinatorial rules, 
reproduces unambiguously  any  unoriented knot diagram. 
A Reed--Muller scheme is used to encode with redundancy 
the eight prototiles into blocks and, once
the blank tile is added and suitably encoded, 
the knot diagram is turned 
into an $N \times N$ mosaic, uniquely associated with a 
string of length $4 N^2 $ bits. The complexity
of the knot, measured topologically by the number of crossings, 
is in turn polynomially related to the number of tiles 
of the associated mosaic, and for knot diagrams of higher complexity
the mosaic encoding provides a design of the knot--based protocol  
which is fault--tolerant under random 1-bit flips. 
It is also argued that the knot mosaic alphabet might be used
in other applications which require high--capacity data transmission.
}

\onecolumn \maketitle \normalsize \vfill

\section{\uppercase{Knot-based protocol: review of the DT code}}
\label{sec:review}
The 
theoretically secure protocol proposed in (Marzuoli \& Palumbo, 2011)
is based on purely topological knot theory. The scheme 
relies on the `easy' problem of associating with prime knots
listed in Knot Tables their Dowker--Thistlethwaite codes, 
numerical sequences which
are different for inequivalent knots. 
Then the `difficulty' of factorizing
complex knots generated by composing prime knots is exploited to
securely encode the given message.
The scheme combines an asymmetric public key protocol
with symmetric private ones and is briefly reviewed in the Appendix. 
In the following two sections the protocol will be improved 
for what concerns the efficiency of the
encoding of  knot diagrams and its  error robustness. 

In order to explain the DT coding used in the original
proposal for the protocol, a few basic notions of topological knot theory 
have to be recalled (Rolfsen, 1976).
A knot $K$ is a continuous embedding of the circle $S^1$ 
into the Euclidean $3$--space $\mathbb{R}^3$. Knots can be oriented or unoriented,
and collections of a finite number of interlaced knots are called links
(in the cryptographic protocol only knots will be used).
Referring for simplicity to the unoriented case, two knots $K_1$ and $K_2$ are said to be 
equivalent, $K_1 \sim K_2$, if and only if they are (ambient) isotopic. An isotopy
is a continuous deformation of the shape of, say, $\,K_2 \subset \mathbb{R}^ 3$
which makes $K_2$ identical to $K_1$ without cutting and gluing back the `closed string' 
$K_2$. 

The diagram of a knot $K$ is its projection 
on a plane $\mathbb{R}^2 \subset \mathbb{R}^3$, in such a way that no point belongs to
the projection of three segments, namely the singular points in the diagram are only
transverse double points. Such a projection, together with `over' and `under' information
at the crossing points --depicted in figures by breaks in the under--passing segments--
is denoted by the same symbol $K$.
In Knots Tables (see (Hoste et al., 1998)
and the \textit{Knot Atlas} on Wikipedia) standard ({\em i.e.}  associated
with minimal projections)
diagrams of unoriented `prime' knots are listed by increasing crossing numbers
as $\chi_n$, where $\chi$ is the number of crossings and
$n= 1,2,\dots$ enumerates in a conventional way the 
knots with the same  $\chi$. The  `unknot' or trivial knot $K_{\bigcirc}$
is such that $\chi (K_{\bigcirc})=0$ and its standard projection is the  circle.
Recall that a prime knot is a non--trivial knot which 
cannot be decomposed into two or more non--trivial knots. 
Decomposition is  the inverse of the topological
operation of `composition' of knot diagrams.
More precisely, given two knot diagrams $K_1$ and $K_2$, it is possible to
draw a new knot  by removing a small segment from each knot and
then joining  the four endpoints by two new arcs.
The resulting diagram is their  connected sum,
denoted by $K_1$ $\#$ $K_2$. Below it is shown the connected sum
of  the trefoil knot $K_1$ (configuration $3_1$ in Knot Tables)
with its mirror image $K_2$, giving rise
to the 6-crossing `granny' knot.
\[
\renewcommand{\labelstyle}{\scriptstyle}
\begin{xy}
0;/r8mm/:
,{\xcapv-|{\,}}
, +(0,1) ,
{\xcaph|{\,}
\xunderh|{\,}%
}
, -(2,1),{\xoverh|{\,}}
,-(1,1),{\xcapv-|{\,}\xcaph-|{\,}}
, +(0,1),{\xunderh|{\,}}
, +(0,1),{\xcapv|{\,}}
, +(0,2) , \xcaph|{\,}
, +(0,-1)  , 
{\xcapv-|{\,} 
\xoverh|{\,}}
, +(-2,-1) \xcaph-|{\,}
,+ (1,0),{\xcaph-|{\,}}
,+(0,1), {\xcapv|{\,}}
, +(-1,2) ,
\xunderh|{\;\;\;\;\;\;K_1 \, \mathbf{\#} \,K_2} 
, +(0,1) \xcapv|{\,}
, +(-1,1) \xcaph|{\,} 
,-(2,0) \xoverh|{\,}
\end{xy}
\]
The Dowker--Thistlethwaite (DT) notation (or code) is defined
for oriented knots and assigns to each 
planar diagram its (minimal) DT sequence.
Given for instance an oriented alternating knot with $\chi$
crossings (namely a diagram with an alternating sequence 
of over and under--crossings) the associated 
DT sequence is built iteratively:  i) start 
labeling an arbitrarily chosen crossing with  1; ii) then, following
the given orientation,
go down the strand to the next crossing and denote it by 2;
iii) continue around the knot  until each
crossing has been numbered twice. Thus each crossing 
is decorated with a pair of even/odd positive numbers, 
running from 1 to 2$\chi$, as shown below for the knot
listed as $5_1$. 
\[\xy
(6,9)*{}="1";
(-8.5,-1)*{}="2";
"1";"2"**\crv{~*=<.5pt>{.}
(0,30)}?(.75)*\dir{>}+(-2,2)*{3}+(-5,-3)*{8};
(-6.5,8)*{}="1";
(-.5,-9)*{}="2";
"1";"2"**\crv{~*=<.5pt>{.}
(-28.5,9.3)}?(.7)*\dir{>}+(-2,-2)*{9}+(2,-3)*{4};
(-9.5,-3.35)*{}="1";
(8.5,-3)*{}="2";
"1";"2"**\crv{~*=<.5pt>{.}
(-17.67,-24.19)}?(.7)*\dir{>}+(-1,-3)*{5}+(6,1)*{10};
(1,-10)*{}="1";
(6.5,7.13)*{}="2";
"1";"2"**\crv{~*=<.5pt>{.}
(17.67,-24.19)}?(.7)*\dir{>}+(3,-1)*{1}+(1,7)*{6};
(11,-1)*{}="1";
(-4,8)*{}="2";
"1";"2"**\crv{~*=<.5pt>{.}
(28.5,9.3)}?(.93)*\dir{>}+(1,3)*{7}+(6,3)*{2};
\endxy\]
For generic, non--alternating prime knots 
(which actually appear in tables for crossing numbers greater than
7), the DT coding is slightly modified by making 
the sign of the even numbers positive
if the crossing is on the top strand, and 
negative  if it is on the bottom strand. 
Since any  sequence is dependent on both a minimal
projection and  the choice of a starting point, the mapping
between knots and their DT sequences is in general one--to--many. 
In the following section a new type of encoding which overcome
these ambiguities is proposed.

\section{\uppercase{Knot mosaics and their block encoding}}
\label{sec:mosaics}

Kauffman and Lomonaco  introduced in (Lomonaco \& Kauffman, 2008) a {\em  mosaic system}
--made of eleven elementary building blocks-- with the aim of addressing what they call 
`quantum knots', namely quantum observables arising in the framework of 3-dimensional
topological quantum field theories. The prototiles we are going to employ here are a subset
of Lomonaco--Kauffman mosaics which suffices to reconstruct the cores of diagrams of (prime or
composite) knots on the basis of purely combinatorial rules to  be 
addressed in the following section. 

\begin{figure}[!h]
  \centering
   {\includegraphics[width=7.5cm]{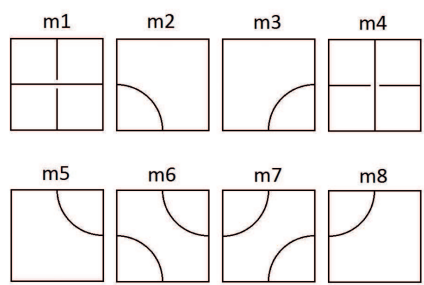}}
  \caption{The eight knot prototiles.}
  \label{fig:tiles2}
 \end{figure}

Let $\mathcal{M}_{(8)}$ denotes the ordered set of  eight knot mosaics (prototiles)
$m_1,$ $m_2,$ 
$m_3,$ $m_4,$
$m_5,$ $m_6,$
$m_7,$ $m_8$
depicted in Fig. \ref{fig:tiles2}. Similarly to what happens in tiling  (a portion of) a plane
with a given set of prototiles, the single mosaics will be 
assembled as they stand (neither rotation nor reflection allowed). Note however
that, unlike most
commonly used tiling prescriptions, the set that is being used here is closed under rotations
of the single mosaics (the most economical set would  include just $m_1,$ $m_2,$ $m_6$,
but then rotations should be allowed). 

These objects can  be associated with digital sequences through
encoding maps $\mathcal{M}_{(8)} \rightarrow \{0,1\}^{*}$ that
can be chosen  in many different ways
(Pless,  1982). For instance the code space
$\Sigma_{(3)} := \{0,1\}^{3}$ of 3-bit sequences would provide 
a very simple (but not fault--tolerant)
one--to--one encoding of the eight mosaics. 
Variable--length codes do not seem particularly
suitable in the present context, where each mosaic is
in principle on the same footing as each other,
and the encoding of long sequences of tiles
would become overwhelming.
To achieve a sufficient degree of  redundancy  
(the basic requirement of any  fault--tolerant coding protocol)
still keeping a block design, consider the injective  map
\begin{equation}\label{encmap1}
\mathcal{E} \;:\; \mathcal{M}_{(8)} \,\rightarrow \; \Sigma_{(4)} := \{0,1\}^{4}
\end{equation}
defined by the correspondences  given in  Table \ref{tab:8mos}.

\begin{table}[h]
\caption{Encoding of the eight knot mosaics into 4-bits strings} \label{tab:8mos}  \centering
\vspace{.5 cm}
\begin{tabular}{|c|c|c|c|}
\hline
$m_1$ & $m_2$ & $m_3$ & $m_4$  \\
\hline
\,$0000$\, & \,$0101$\, & \,$1010$\, & \,$1111$\,  \\
\hline
\hline
$m_5$ & $m_6$ & $m_7$ & $m_8$ \\
\hline
\,$0011$\, & \,$0110$\, & \,$1001$\, & \,$1100$\, \\
\hline
\end{tabular}
\end{table}
The valid words in the code space $\Sigma_{(4)}$ are a subset characterized by
the property of being identical to the eight sequences of
the binary Reed--Muller code $\mathcal{R}(1,2)$. Mutual Hamming distances
between these codewords are easily evaluated: $d(m_1,m_4)=$ 
$d(m_2,m_3)=$ $d(m_5,m_6)=$ $d(m_6,m_7)= 4$, and all others are equal to $2$.
The minimal distance associated with the encoding (\ref{encmap1}) is
\begin{equation}\label{encmap2}
d_{\,\text{min}} \;[\mathcal{E}\,(\mathcal{M}_{(8)})\, \subset \,
\Sigma_{(4)}]\,= \,2,
\end{equation}
thus providing a random error detection ability equal to $1$ (a single
bit--flip cannot turn one codeword into another). 

In order to carry out the assembling of  knot mosaics, the 
empty prototile (blank tile) has to be added and suitably represented in
the codespace $\Sigma_{(4)}$.
Actually, as will be described in the following section, a single blank
tile does not suffice to achieve redundancy, but rather four empty mosaics
endowed with double arrows are needed. 
Denoting by $\mathcal{B}_{(4)}$ the ordered set of the four mosaics
$b_1,$ $b_2,$ $b_3,$ $b_4$, the domain of the encoding map is extended to
\begin{equation}\label{encmap3}
\mathcal{E} \;:\; \mathcal{B}_{(4)} \cup \mathcal{M}_{(8)} \,\rightarrow \; \Sigma_{(4)},
\end{equation}
where the $b$'s are associated with previously unassigned codewords
according to the list in Table \ref{tab:4bla}. 

\begin{table}[h]
\caption{Encoding of the four blank mosaics} \label{tab:4bla}  \centering
\vspace{.5 cm}
\begin{tabular}{|c|c|c|c|}
\hline
$b_1$ & $b_2$ & $b_3$ & $b_4$  \\
\hline
$1000$ & $0010$ & $0100$ & $0001$  \\
\hline
\, & \, & \, & \,  \\
$\boxed{^{\rightarrow} \hspace{- .1 cm} _{\downarrow}} $ &
$\boxed{_{\leftarrow} \hspace{- .1 cm}  ^{\downarrow}}$ &
$\boxed{^{\uparrow} \hspace{- .1 cm}  _{\leftarrow}}$ &
$\boxed{_{\uparrow} \hspace{- .1 cm} ^{\rightarrow}}$ \\
\, & \, & \, & \,  \\
\hline
\end{tabular}
\end{table}
It can be easily checked that the distances among  $b$'s are
all equal to $2$, namely
\begin{equation}\label{encmap4}
d_{\text{min}} \;[\mathcal{E}\,(\mathcal{B}_{(4)})\, \subset \,
\Sigma_{(4)}]\,= \,2,
\end{equation}
while distances between a mosaic $m_i$ $(i=1,2,\dots,8)$
and a blank tile $b_{\alpha}$ $(\alpha=1,2,3,4)$
amount to either $1$ or $3$. The particular correspondences in 
Table \ref{tab:4bla} are chosen in order to maximize
distances to particular types and pairings of $m$'s
which are placed in the boundary layer of an $N \times N$
mosaic ({\em cfr.} diagrams in the following section).

\section{\uppercase{Encoding process of diagrams into $N \times N$ mosaics }}
\label{sec:NNmosaics}

\begin{figure}[!ht]
  \centering
   {\includegraphics[width=6.5cm]{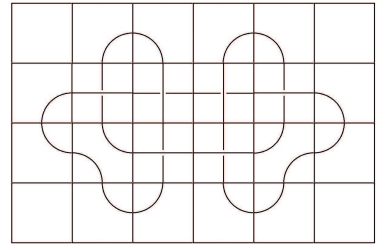}}
  \caption{The mosaic of the granny knot.}
  \label{fig:tiles6}
 \end{figure}

As a first instance of what is meant by an assembled knot diagram
made out of $m$ and $b$ tiles, in Fig. \ref{fig:tiles6} the composite
6-crossing granny knot already shown in section \ref{sec:review} is depicted.

\begin{figure}[!ht]
  \centering
   {\includegraphics[width=5.5cm]{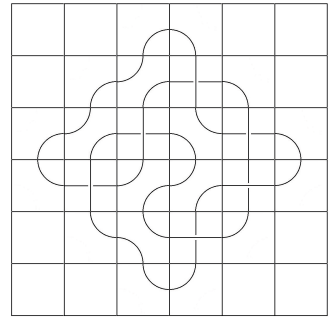}}
  \caption{The mosaic of the $6_1$ knot.}
  \label{fig:tiles7}
 \end{figure}

\begin{figure}[!ht]
  \centering
   {\includegraphics[width=5.5cm]{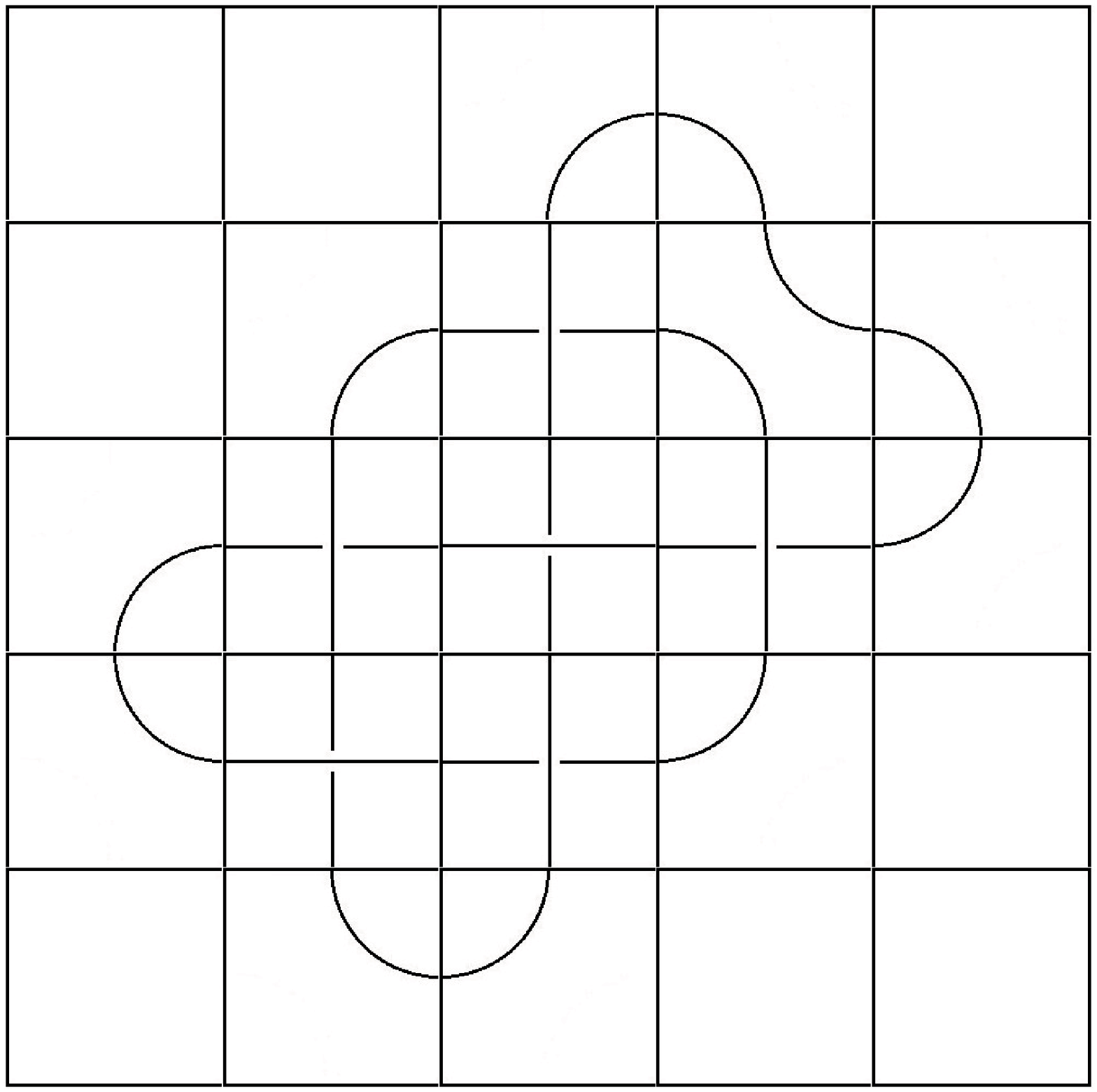}}
  \caption{The mosaic of the $6_2$ knot.}
  \label{fig:tiles4}
 \end{figure}

\begin{figure}[!ht]
  \centering
   {\includegraphics[width=5.5cm]{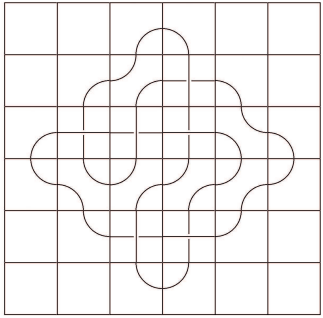}}
  \caption{The mosaic of the $6_3$ knot.}
  \label{fig:tiles3}
 \end{figure}

This rectangular $6 \times 4$ mosaic can be completed with two more rows of blank tiles
to get a $6 \times 6$ square and, as can be easily checked, also all 6-crossing 
{\em prime} knots can be arranged into such a square by suitably combining
the $m$-type and blank tiles, see Fig. 
\ref{fig:tiles7}, \ref{fig:tiles4}, \ref {fig:tiles3}.
(recall that there are exactly  three prime
knots with 6 crossings, listed as $6_1$, $6_2$, $6_3$ in Knot Tables).

While the above examples can be handled quite easily, in  case
of a generic diagram of some prime knot $K$ with $\chi \, (K)$ crossings,  
global combinatorial  rules  are needed to reconstruct (and coding) the associated
mosaic. Note first that the extension of the square (rectangle), not known {\em a priori}, 
can be evaluated since any tile containing a crossing ($m_1, m_4$ of Fig.
\ref{fig:tiles2}) is topologically interconnected to eight  tiles surrounding it. 
Then an upper bound on the size of an $N \times M$  mosaic is given by 
\begin{equation}\label{NNCr}
N \times M \;\leq \; 8\chi \,,
\end{equation}
so that  the operation of converting any  standard knot diagram into a  mosaic 
can be efficiently performed. 

\begin{figure}[!ht]
  \centering
   {\includegraphics[width=6.5cm]{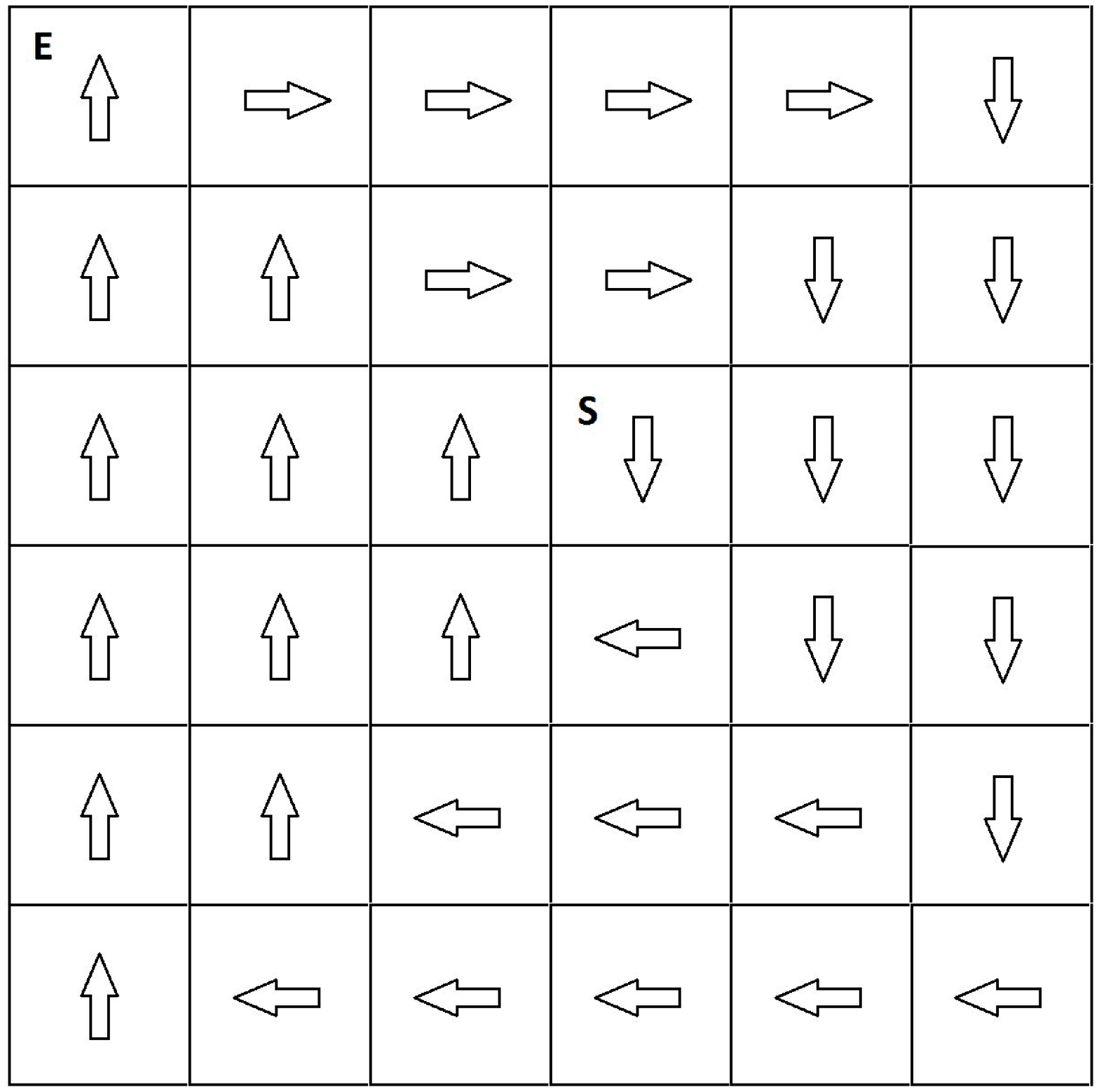}}
  \caption{The clockwise spiral
path starting in {\bf S} and ending in {\bf E}
for assembling prototiles of the 6-crossing knot diagrams into  $6 \times 6$
mosaics.}
  \label{fig:arrows}
 \end{figure}

Focusing on square mosaics, since their assembling must proceed with no reference to the
size of the resulting table, it is worth starting from
an internal tile and following a spiral path, moving {\em e.g.} in the clockwise
direction, as shown in Fig. \ref{fig:arrows} for a $6 \times 6$ mosaic. 

As for blank
tiles, placed at the corners of $ 6 \times 6$ mosaics in Fig. \ref{fig:tiles7}, \ref{fig:tiles4},
\ref{fig:tiles3}, their decorations with double arrows introduced in 
Table \ref{tab:4bla}
can now be explained. Thus $b_1$ must be used to fill the empty
squares around the upper right corner, $b_2$ around the lower right,
$b_3$ around the lower left, and $b_4$ around the upper left.
Referring, {\em e.g.}, to  $b_1$ in connection with Fig. \ref{fig:tiles7},
both arrows of tile $b_1$ are activated in the empty square 
located at the upper right corner;
the rightward  arrow agrees with the spiral pathway one mosaic left 
to the corner and the downward  arrow agrees with the spiral pathway 
one mosaic below the corner.

Before going through an analysis of topological prescriptions
which will further enforce the coding procedure with respect to
fault tolerance, it is necessary to extend the encoding map (\ref{encmap3})
to deal with $N \times N$ mosaics. Denoting by a $^*$
the set of all words based on the two alphabets ($12$ knot tiles and 
$16$ 4-bit sequences) define  
\begin{equation}\label{extemap}
\mathcal{E}^{(N \times N)} \;:\; (\,\mathcal{B}_{(4)} \cup \mathcal{M}_{(8)}\,)^* 
\,\rightarrow \; \Sigma_{(4)}^*,
\end{equation}
where accepted codewords (of length $4N^2$ bits)
are those associated with
knot mosaics that can be arranged into the given square.

The topological (combinatorial) prescriptions that have to
be taken into account  traveling along the spiral encoding 
path of a given  $N \times N$
knot mosaic are summarized as follows.
\begin{itemize}
\item[i)] The knot  is a continuous closed path
(over and under--crossing points are artifacts
due to the fact that we deal with a planar projection)
and thus each mosaics added to the previous one must match 
correctly the knot strand. In other words, any
arrangement of mosaics giving rise at some point
to disconnected arcs or lines is forbidden.
\item[ii)]  Most critical situations may occur at the crossings points
(knot mosaics $m_1$ and $m_4$) because 
an accidental swap would change the topology of the knot.
The encoding given in section \ref{sec:mosaics}, Table
\ref{tab:8mos}, is such that $d(m_1,m_4)=4$ and thus
this type of error is actually highly suppressed.
\item[iii)] Away from  crossing points, but still
in the core region of the mosaic, another kind of 
mismatching can occur whenever a double--arc tile
($m_6,m_7$) is turned accidentally into a single-arc one
($m_2,m_3,$ $m_5,m_8$): then either a discontinuity
(see prescription i)) or  an improper
closure of the knot string is created. In the latter case 
the resulting configuration would correspond to a link,
namely a multicomponent knot, contrary to the basic assumption
which must hold true in the knot--based cryptographic protocol.
\item[iv)] Finally, going through the most external
layer of the mosaic, accidental swaps
could occur between single--arc tiles, between the latter and blank tiles
and between blank tiles.
In all these situations disconnected patterns must be ruled out
again on the basis of i) (recall also that the minimum distances among the $m$'s
and among the $b$'s is 2). The encoding prescription for 
the blank tiles (Table \ref{tab:4bla})
together with the observation that $m$-tiles are arranged in
pairs in the boundary layer ($m_3$-$m_2$, $m_2$-$m_8$
$m_8$-$m_5$, $m_5$-$m_3$) ensure that distances between
anyone of these $m$ and the two admissible contiguous blank tiles
are equal to 2. 
\end{itemize}

\section{\uppercase{Conclusions and outlook}}
\label{sec:conclusion}

It has been shown that the mosaic encoding of the
knot--based cryptographic protocol proposed in (Marzuoli \& Palumbo, 2011)
is efficient (with respect to increasing  complexity of  knot
diagrams) and robust against random 1-bit flips of
the encoded string of $4N^2$ bits. Such procedure
can be applied to both prime knots in standard Knot Tables
and composite knots used in the protocol.
The rules described in the previous section, in particular
the spiral pathway in the $N \times N$ mosaic
and  the connection property i), enforce fault--tolerance 
since are related to topological, global features:
most probable errors
would turn connected knot diagrams into  
collections of tiles still joined together but not
corresponding to a connected knotted curve.

The system of knot mosaics, introduced
in (Lomonaco \& Kauffman, 2008) and used here for specific
encoding purposes, might be
employed in at least two more contexts. The first one
is related to the fact that they are `prototiles', so that it would
be  interesting to ask whether  
they constitute a Wang set, namely if they are able to generate
aperiodic tiling of the plane (Grunbaum \& Shepard, 1987). It does
not seems so, but work is in progress to improve (or disprove) this conjecture,
which could  have interesting consequences also for
open problems in topological knot theory.
Recall that  Wang tiles have not only a foundational interest in logic
and in information theory, but are used also in
applied computer science, see {\em e.g.} (Cohen et al., 2003)
in connection with image and texture generation.

A second remark resorts to the observation that 
the mosaic system is actually an alphabet, as 
the mapping (\ref{extemap}) suggests. Large knot mosaics,
with varying $N$ and embedded knotted curves, might
be used, in turn, to modeling and encoding large
sets of data. Once shown
that the procedure described in the previous section
provides a sort of `topologically--protected' encoding,
it can be argued that such further extensions
could define a new efficient method for data transmission.

\vfill
\bibliographystyle{apalike}
{\small
\bibliography{example}}
\begin{enumerate}
\item Cohen, M. F., Shade, J., Hiller, S. and Deussen, O. 2003
`Wang tiles for image and texture generation',
\textit{ACM Transactions on Graphics}, 
vol. 22, n. 3, pp. 287-294.
\item Grunbaum, B and Shepard, G. C. 1987,
\textit{Tilings and patterns}, W. H. Freeman and Company.
\item Hoste,  J. Thistlethwaite, M. and Weeks, J.
1998 `The first 1,701,935 knots'
\textit{Math. Intelligencer},  vol. 20, n. 4, pp. 33-48.
\item Lomonaco, S.J. and Kauffman, L. H. 2008
`Quantum knots and mosaics', 
\textit{Quantum Inf. Process.}, vol 7, pp.85-115.
\item  Marzuoli, A. and   Palumbo, G. 2011,
`Post quantum cryptography from mutant prime knots'
\textit{Int. J. Geom. Methods in Mod. Phys.}, vol.8, n. 7, pp. 1571-1581.
\item Pless, V. 1982, \textit{Introduction to the Theory of 
Error-Correcting Codes}, Wiley-Interscience Series in Discrete Mathematics,
John Wiley \& Sons, New York.
\item Rolfsen, D. 1976, \textit{Knots and Links},
Publish or Perish, Berkeley, CA.
\end{enumerate}

\section*{\uppercase{Appendix}}

As is well known most RSA--type protocols are based on the  
computational complexity of  factorization of prime numbers, 
because the generators are two large prime numbers (p
and q) and the public key  is the product of them (N = pq).
Once given N, decrypting the message needs the knowledge 
of its prime factors, and
this is of course a computationally hard problem. Note however that
public key algorithms are very costly in terms of
computational resources. The time it takes the message to be encoded
and decoded is relatively high and  this is actually
the main drawback of (any)
asymmetric decoding. This problem can be overcome or even solved
by using a symmetric key together with the asymmetric one,
as done in (Marzuoli \& Palumbo, 2011) for the
knot--based cryptosystem. The following brief review refers to
the original formulation based on the DT coding. The translation
in terms of the (most reliable and falt--tolerant) 
mosaic encoding proposed in this paper would be straightforward.

A sender $\mathcal{A}$ must prepare a secret message for the receiver
$\mathcal{B}$ and they share
the same finite list of prime knots $K$'s. 
The message $\mathbf{M}$ will be  built by resorting to
a finite sequence of (not necessarily prime) knots $L_{1},...,L_{N}$ 
according to the following steps.

\begin{itemize}
\item[I)] Through a standard RSA  protocol, 
$\mathcal{B}$  sends to $\mathcal{A}$ an ordered  sublist of $N$
prime knots taken from current available Knot Tables, $K_{1},...,K_{N}$,
together with mutation instructions to be applied to each $K_{i}$.
(The operation called `mutation' amounts to remove a portion of
the knot diagram with four external legs,  replace it 
with the configuration obtained by rotating the original
one, and then gluing back the tips of the strands.)

A second list $K'_{1},...,K'_{N}$ is generated by
picking up definite mutations of the original sequence. 
\item[II)]  $\mathcal{A}$ takes $K'_{1},...,K'_{N}$ 
 and performs a series of ordered connected sums
$$ L_{1} \# K'_{1},\, L_{2} \# K'_{2},\, \ldots, L_{N} \# K'_{N}
$$
with the knots $L_{1},...,L_{N}$ associated with the message to be sent.\\
These composite  knots are now translated (efficiently) into 
Dowker--Thistlethwaite sequences and sent to $\mathcal{B}$ . Obviously 
at this stage everyone has access to these 
strings of relative integers.
   
\item[III)] $\mathcal{B}$  receives the (string of) composite knots. 
Since he knows the DT sub--codes
for the prime knots of the shared list, 
he can decompose the composite knots, thus
obtaining the DT code for every $L_{i}$. 
Then the planar diagrams of 
$L_{1},...,L_{N}$ can be uniquely recovered. 
\end{itemize}

Basically we are using in the protocol both a public key (step
I) and a private key (step II). In fact the message is encrypted (by
$\mathcal{A}$) and decrypted (by $\mathcal{B}$) using the same key,  the sequence of prime
knots that they share (secretly) thanks to  step I).

\vfill
\end{document}